\documentclass[preprint,review,12pt]{elsarticle}
\usepackage{amsmath,amssymb,amsfonts}
\usepackage{color,bm}
\usepackage{hyperref}
\usepackage{zref-clever}
\usepackage[margin=2cm]{geometry}
\usepackage{ifthen}
\usepackage{lineno}
\usepackage{mathtools}
\usepackage{mhchem}
\usepackage[separate-uncertainty=true]{siunitx}
\usepackage{subcaption}
\usepackage{tikz}
\usetikzlibrary{arrows.meta}
\usetikzlibrary{calc}
\usetikzlibrary{decorations.pathreplacing}
\usetikzlibrary{math}
\usetikzlibrary{positioning}

\zcsetup{cap,abbrev}

\graphicspath{{fig/}}

\journal{Elsevier}

\begin{document}

\begin{frontmatter}
    \title{The Fast Stochastic Matching Pursuit for Neutrino and Dark Matter Experiments}

    \author[1,2,3]{Yuyi Wang}
    \author[1,2,3]{Aiqiang Zhang}
    \author[1,2,3]{Yiyang Wu}
    \author[1,2,3]{Benda Xu\corref{cor1}}%
    \ead{orv@tsinghua.edu.cn}
    \author[1,2,3]{Xuewei Liu}
    \author[4]{Jiajie Chen}
    \author[1,2,3]{Zhe Wang}
    \author[1,2,3]{Shaomin Chen}

    \cortext[cor1]{Corresponding author}
    \affiliation[1]{
        organization={Key Laboratory of Particle \& Radiation Imaging (Tsinghua University)},
        addressline={Ministry of Education},
        country={China}}
    \affiliation[2]{
        organization={Department of Engineering Physics},
        addressline={Tsinghua University},
        cite={Beijing},
        postcode={100084},
        country={China}}
    \affiliation[3]{
        organization={Center for High Energy Physics},
        addressline={Tsinghua University},
        cite={Beijing},
        postcode={100084},
        country={China}}
    \affiliation[4]{
        organization={Department of Computer Science and Technology},
        addressline={Tsinghua University},
        cite={Beijing},
        postcode={100084},
        country={China}}


\begin{abstract}
    Photomultiplier tubes~(PMTs) are widely deployed at neutrino and dark
    matter experiments for photon counting.  When multiple photons hit a
    PMT consecutively, their photo-electron~(PE) pulses pile up to hinder
    the precise measurements of the count and timings.  We introduce Fast
    Stochastic Matching Pursuit~(FSMP) to analyze the PMT signal
    waveforms into individual PEs with the strategy of reversible-jump
    Markov-chain Monte Carlo.  We demonstrate that FSMP improves the
    energy and time resolution of PMT-based experiments and gains
    acceleration on GPUs. It is suitable for dynode PMTs, and is extensible to microchannel-plate~(MCP)
    PMTs.  In the condition of our laboratory
    characterization of 8-inch MCP-PMTs, FSMP improves the energy
    resolution by up to \SI{10}{\percent} from the conventional method of
    waveform integration.
\end{abstract}


\end{frontmatter}


\section{Introduction}
Large liquid detectors equipped with photomultiplier tubes~(PMT) are widely used for detecting the invisible, enigmatic, and challenging-to-detect neutrinos and dark matter.
The electronic systems read photon-induced pulses embedded in the time series of PMTs' voltage outputs, or \emph{waveforms}.  Experiments deploying full waveform readout include KamLAND~\cite{kamland_collaboration_reactor_2013}, IceCube~\cite{stokstad2005design}, Borexino~\cite{gatti_borexino_2001}, JUNO~\cite{Petitjean_2022,anfimov_large_2017}, Jinping Neutrino Experiment~(JNE)~\cite{beacom_physics_2017,xu_jinping_2020,xu_design_2022}, as well as XMASS~\cite{abe_measurement_2018}, DEAP-3600~\cite{AMAUDRUZ20191}, and PandaX-4T~\cite{Yang_2022}.

\label{sec:intro-lc}
After a scintillation photon is emitted in an event and reaches a PMT, it may strike the photocathode and induce the photoelectric effect. A photoelectron (PE) may be released into the vacuum area of the PMT. Sometimes, multiple PEs can be produced from a single photon when its energy is greater than or equal to twice the work function of the material~\cite{faham_measurements_2015,lopez_paredes_response_2018}, but they could be conceptually treated as one PE with multiple charges. Without loss of generality, we can consider each photon producing at most one PE in this context.
The PEs follow an \emph{inhomogeneous} or \emph{time-dependent} Poisson point process~\cite{4324923,CRESSIE199493,baddeley_spatial_2016} with intensity function $\lambda\phi(t-t_0)$, where $\lambda$ is the expectation of the number $N$ of PEs, $\phi(t-t_0)$ is the probability density, and $t_0$ is the event time.
It could be calibrated by photon counting methods~\cite{birks2013theory,guo_slow_2019,OConner1984-cr}.

For simplicity, we assume that the \emph{electronic noise} is white that follows a normal distribution independent of time: $\epsilon(t)\sim\mathcal{N}(0,\sigma_\epsilon^2)$~\cite{lindgren2013stationary}. An electrical current is formed when electrons impact the anode. This
current is converted typically through a transimpedance amplifier to a voltage pulse for waveform readout~\cite{jetter_pmt_2012}, and the
resulting \emph{charge} can be calculated via waveform integration~\cite{huang_flash_2018}. This article uses \si{pC} as the unit of charge.

\label{sec:intro-ser}
Daya Bay~\cite{jetter_pmt_2012} and JUNO~\cite{zhang_comparison_2019} describe the pulse shape from a single PE as being log-normally distributed.
Luo~et~al.~\cite{luo_reconstruction_2023} and Zhang~et~al.~\cite{zhang_performance_2023} use the ex-Gaussian distribution.
Knoll~\cite{knoll_radiation_2010} uses mixed exponential functions derived from the parallel RC circuit.
There is no established unique model; we shall use the ex-Gaussian distribution without loss of generality.
The waveform of a PMT of PE $i$ is $q_i V_\mathrm{PE}(t-t_i)$, where $q_i$ is the PE charge, $t_i$ is the PE time, and $V_{\mathrm{PE}}$ is the normalized shape of the single electron response (SER). With $V_{\mathrm{PE}}$ and the electronic noise $\epsilon(t)$, the final waveform $w(t)$ is
\begin{equation}
    \label{eq:waveform}
    w(t)=\sum_{i=1}^N q_i V_\mathrm{PE}(t-t_i)+\epsilon(t).
\end{equation}

To reconstruct the energy and time of the events from the waveforms, a common method is to utilize the integrated charge (also known as the \emph{charge method})~\cite{zhang_comparison_2019} as a predictor of \emph{visible energy}, and to locate the peaks of the waveforms, measuring the 10\%-rising-edge~\cite{zhang_performance_2023,knoll_radiation_2010} as PE times.
Matched filters~\cite{1057571} can be used to suppress electronic noises and improve the signal-to-noise ratio; however, they are insufficient for retrieving the times and number of PEs. The filtered signal might be worse at distinguishing the PEs that are very close.
More sophisticated approaches use fitting or deconvolution~\cite{zhang_comparison_2019,luo_reconstruction_2023,knoll_radiation_2010,xu_towards_2022} with empirical single PE templates to obtain the charge and PE arrival times together.

When the time difference between two PEs is small, their waveforms \emph{pile up}~\cite{luo_pulse_2018}, which prevents reliable counting of the PEs. A possible approach is to utilize the charge directly~\cite{grassi_charge_2018}, while it loses time information in the waveforms and makes the energy resolution worse.
Luo~et~al. propose a waveform fitting method assuming two pulses~\cite{luo_pulse_2018}; Akashi-Ronquest~et~al. try to better find the most probable count of PEs based on the Bayesian theorem and charge method~\cite{AKASHIRONQUEST201540}. Both retrieve a fixed number of PEs and prevent the estimation of uncertainty.
Huang~et~al. analyze waveforms with a fitting method based on a single PE waveform, but find that the failure rate increases with an increasing count of PEs~\cite{huang_flash_2018}.
Grassi~et~al. try another fitting method independently, finding that it becomes slow and unreliable with large pile-ups~\cite{grassi_charge_2018}.
Jiang~et~al.~\cite{jiang2024machinelearningbasedphotoncounting} apply machine learning to photon counting by formulating it as a classification task, with a hyperparameter limiting the maximum resolvable PE count. Consequently, the method fails for waveforms containing more than nine PEs.
Ideally, waveform analysis should find all possible PE times and counts from a trans-dimensional space, and retrieves the most possible solutions. It is a hierarchical and discrete-continuous challenge.
\emph{Fast Stochastic Matching Pursuit}~(FSMP) is a fast and flexible algorithm that enables jumping between different dimensions and utilizes all information from the waveforms.  It was introduced in our previous publication~\cite{xu_towards_2022}, and then utilized to analyze a variety of PMTs, and most notably adopted to the new microchannel-plate~(MCP) PMTs~\cite{zhang_performance_2023}, showing outstanding performance.

JNE, a large scintillator detector currently under
construction, serves as the context for this
discussion~\cite{xu_design_2022}.  This experiment aims to detect
neutrinoless double beta decays, geo-neutrinos, and solar neutrinos,
which necessitates high energy and time resolutions for particle
identification~\cite{PhysRevC.107.014323} and distinguishing Cherenkov
and scintillation
signals~\cite{aberle_measuring_2014,elagin_separating_2017,yoshiyuki_direct_2021,luo_reconstruction_2023}.
To facilitate this, a slow liquid scintillator based on linear alkyl
benzene~(LAB) has been developed~\cite{LI2016303,guo_slow_2019}.
Similar recipes for comparable physical
motivations have been explored by many groups~\cite{biller_slow_2020,dunger_slow-fluor_2022,steiger_development_2024}.
To mitigate signal pile-up from Cherenkov and scintillation light, FSMP is required to resolve the arrival times of individual PEs with a temporal resolution better than $\SI{1}{\nano\second}$, even when their temporal separation is less than $\SI{10}{\nano\second}$.
\zcref{sec:fsmp} gives an introduction to our methodology for tackling the challenge of PE pile-ups.  Performance evaluation based on simulation in \zcref{sec:perf} demonstrates the GPU acceleration and substantial improvements in energy and time resolution.  Application of FSMP to experimental data in \zcref{sec:test} provides an analysis basis to unveil the physics process inside 8-inch MCP-PMTs for JNE.


\section{Methodology}
\label{sec:fsmp}
FSMP uses reversible jump MCMC~(RJMCMC)~\cite{green_reversible_nodate} to analyze the waveforms by sampling from the posterior distribution of PE sequences.

\subsection{Charge model of PMTs}
\label{sec:intro-charge-model}
\begin{figure}[!htbp]
    \centering
    \begin{subfigure}{0.49\linewidth}
        \centering
        \includegraphics[width=0.6\linewidth]{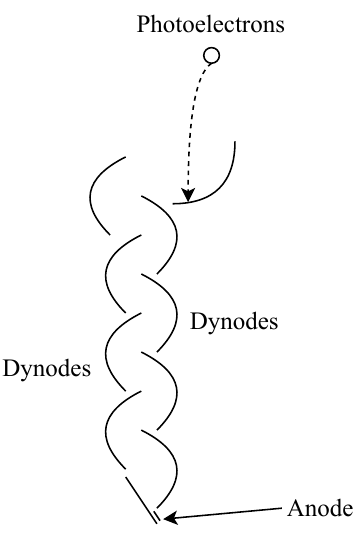}
        \caption{An illustration of a linear-focused dynode PMT~\cite{noauthor_photomultiplier_nodate}.}
        \label{fig:dynode}
    \end{subfigure}
    \begin{subfigure}{0.49\linewidth}
        \includegraphics[width=\linewidth,page=2]{Fig1bd_charge_model.pdf}
        \caption{The normalized charge model of a dynode PMT. It will be fitted into a single normal distribution as the input of FSMP.}
        \label{fig:dynode-charge-model}
    \end{subfigure}
    \begin{subfigure}{0.49\linewidth}
        \includegraphics[width=\linewidth]{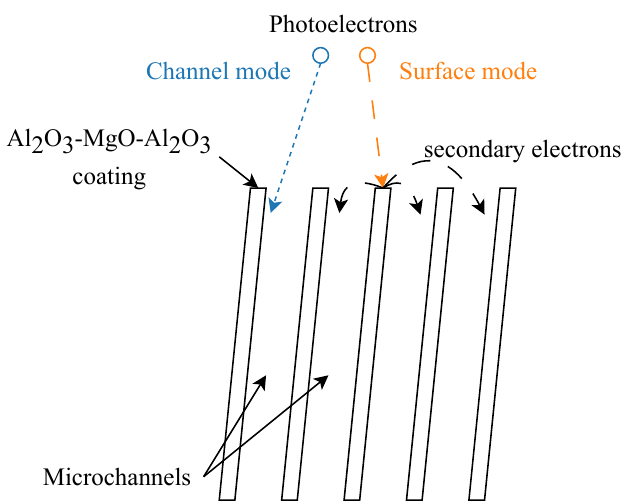}
        \caption{An illustration of MCP and MCP secondary electrons.}
        \label{fig:mcp}
    \end{subfigure}
    \begin{subfigure}{0.49\linewidth}
        \includegraphics[width=\linewidth,page=1]{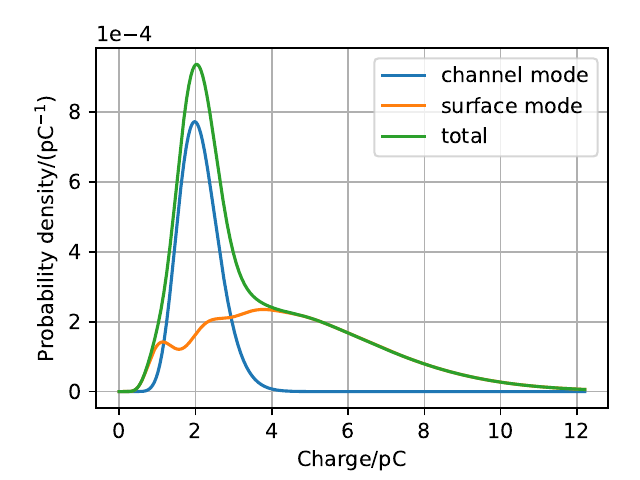}
        \caption{The normalized charge model of an MCP-PMT. It is approximated as a Gaussian mixture to facilitate integration.}
        \label{fig:mcp-charge-model}
    \end{subfigure}
    \caption{The inner structure of dynode and MCP-PMTs, and the charge model of them.}
\end{figure}

As shown in \zcref{fig:dynode}, a dynode PMT multiplies the electrons~\cite{noauthor_photomultiplier_nodate} on each of its many dynodes, and collects them on the anode to produce a signal.
Let $q$ denote the charge of a single PE, following a gamma distribution $\Gamma\left(\frac{\mu_q^2}{\sigma_q^2},\frac{\sigma_q^2}{\mu_q}\right)$~\cite{weng2024single}, shown in \zcref{fig:dynode-charge-model}, approximated as a normal distribution $\mathcal{N}(\mu_q,\sigma_q^2)$ in FSMP. Given that $N$ follows a Poisson distribution $\pi(\lambda)$, the total waveform charge follows a Poisson-gamma compound.
We emphasize that the Poisson expectation $\lambda$, the number of PEs $N$, and the charge of a waveform $Q$ are different quantities. $\frac{Q}{\mu_q}$ is an unbiased estimator of $N$ and $\lambda$, but is inefficient. In \zcref{sec:fsmp-mu}, we will discuss better estimators of $\lambda$ by FSMP.

A new kind of MCP-PMTs used by JNE has high collection efficiency thanks to the atomic layer deposition~(ALD) with \ce{Al2O3}-\ce{MgO}-\ce{Al2O3}, but introduces \emph{jumbo charges}~\cite{weng2024single,zhang_performance_2023}.
In an MCP-PMT shown in \zcref{fig:mcp,fig:mcp-charge-model}, the interaction of a PE with the MCP can occur in two distinct modes~\cite{weng2024single}: it can either enter a microchannel directly (channel mode) or strike the MCP's front surface (surface mode), which subsequently induces the emission of secondary electrons.
The charge of channel mode follows a gamma distribution, while that of surface mode follows a Poisson-gamma compound~\cite{weng2024single}.
Such a charge distribution of the MCP-PMT is parameterized with Gaussian mixtures for the simplicity of the calculations.
The single PE charge model is
\begin{equation}
    \label{eq:mcp-multi-norm}
    \sum_{e} G(e) f_\mathcal{N}(\mu_e,\sigma_e^2)
\end{equation}
where $e$ is the MCP Gaussian index and $G(e)$ is its probability. $G(e)$, $\mu_e$, and $\sigma_e^2$ are the input parameters of FSMP.

\subsection{Formulation}
\label{sec:fsmp-bayesian}
We define the PE sequence $\bm{z}=\{t_1,t_2,...,t_N\}\in \mathcal{T}^N$ as the time of each PE, where $\mathcal{T}$ is the solution space of the times of PEs. For simplicity and to facilitate matrix operations, let $w(t)$ denote $\bm{w}$ interchangeably.
Generally, we need to calculate the total probability $p(\bm{w}|\lambda,t_0)=\sum\limits_{\bm{z}}p(\bm{w}|\bm{z})p(\bm{z}|\lambda,t_0)$.
$p(\bm{w}|\bm{z})=p(\bm{w}|\bm{z},\lambda,t_0)$ is independent of $\lambda$ and $t_0$, because the waveform is determined probabilistically by the PE sequence and the electronic noise.
$p(\bm{z}|\lambda,t_0)$ is defined by the Poisson point process~\cite{baddeley_spatial_2016}
\begin{equation}
    \label{eq:pzmu}
    \begin{aligned}
        p(\bm{z}|\lambda,t_0) \mathrm{d}\bm{z} & = e^{-\lambda}\prod_{k=1}^N \lambda \phi(t_k - t_0) \mathrm{d}t_k     \\
                                               & = e^{-\lambda} \lambda^N \prod_{k=1}^N  \phi(t_k - t_0) \mathrm{d}t_k \\
                                               & = e^{-\lambda} \lambda^N \phi(\bm{z} - t_0) \mathrm{d}\bm{z}
    \end{aligned}
\end{equation}
while $\phi(\bm{z} - t_0)\mathrm{d}\bm{z}$ is an abbreviation for $\prod\limits_{k=1}^N  \phi(t_k - t_0)\mathrm{d}t_k$.
Given $t_0$, we can make a guess $\lambda_0$ and sample $\bm{z}$ from the posterior distribution
\begin{equation}
    \label{eq:pz}
    p(\bm{z}|\bm{w},\lambda_0,t_0)=\frac{p(\bm{w}|\bm{z})p(\bm{z}|\lambda_0,t_0)}{p(\bm{w}|\lambda_0,t_0)}
\end{equation}
and approximate the total probability
\begin{equation}
    \label{eq:reduce-mu0}
    \begin{aligned}
        p(\bm{w}|\lambda,t_0) & =\sum_{\bm{z}}p(\bm{w}|\bm{z})p(\bm{z}|\lambda_0,t_0)\frac{p(\bm{z}|\lambda,t_0)}{p(\bm{z}|\lambda_0,t_0)}               \\
                              & =p(\bm{w}|\lambda_0,t_0)\sum_{\bm{z}}p(\bm{z}|\bm{w},\lambda_0,t_0)\frac{p(\bm{z}|\lambda,t_0)}{p(\bm{z}|\lambda_0,t_0)} \\
                              & =C \mathrm{E}_{\bm{z}}\left[\frac{p(\bm{z}|\lambda,t_0)}{p(\bm{z}|\lambda_0,t_0)}\right]                                 \\
                              & \approx\frac{C}{M}\mathrm{e}^{-(\lambda-\lambda_0)}\sum_{\bm{z}\in\mathcal{Z}}{\left(\frac{\lambda}{\lambda_0}\right)}^N
    \end{aligned}
\end{equation}
where $C$ is a constant, $M$ is the count of sampled $\bm{z}$, and $\mathcal{Z}$ is the set of samples. $\mathrm{E}_{\bm{z}}[\cdot]$ is the expectation by $\bm{z}$, calculated by averaging over samples.

Within a Gibbs sampling framework~\cite{4767596}, the inference of $t_0$ is performed using a Metropolis-Hastings algorithm~\cite{hastings_monte_1970}, while $\bm{z}$ is sampled via RJMCMC~\cite{green_reversible_nodate} to accommodate the unknown number of PEs. The posterior mean is employed as the estimator for $t_0$, a choice that circumvents the bias observed in maximum likelihood estimation~(MLE) of $t_0$ due to the asymmetry of $\phi(t)$.

FSMP is statistically neutral between Bayesian and frequentist approaches. It is theoretically possible to sample $\lambda$ together with $t_0$ and $\bm{z}$.
In the context of event reconstruction utilizing an inhomogeneous Poisson point process, the detector response provides the complete intensity function~\cite{dou_reconstruction_2023,liu_first-principle_2025}, which is parameterized by $\lambda$ and $t_0$ in \zcref{eq:pzmu}. For the scope of this article, our focus is exclusively on waveform analysis; a comprehensive discussion of event reconstruction from waveforms will be presented in a subsequent publication. Given the challenge of establishing a non-informative standalone prior, $\lambda$ is inferred through MLE:
\begin{equation}
    \hat{\lambda}_\mathrm{FSMP}=\underset{\lambda}{\arg \max}~p(\bm{w}|\lambda,t_0)
\end{equation}

\subsection{Preparation of the solution space}
\label{sec:fsmp-lucyddm}
By the Perron–Frobenius theorem~\cite{kifer1996perron}, the close-to-truth initial states of $\bm{z}$ and $t_0$ make the Markov chain of FSMP converge faster, provided that it is irreducible and aperiodic.

We aim to estimate $q(t) = \sum\limits_i q_i \delta(t-t_i)$ through deconvolution, to provide a solution space and an initial guess for $\bm{z}$.
Ignoring the electronic noise, the waveform in \zcref{eq:waveform} is expressed as a convolution
\begin{equation}
    w(t)=q\otimes V_\mathrm{PE}
\end{equation}
So the estimation $\hat{q}(t)$ could be calculated as $\bm{w}\oslash V_\mathrm{PE}$, where $\oslash$ represents deconvolution.
An iterative deconvolution algorithm developed by Lucy~\cite{lucy_iterative_1974} is suitable for the nonnegative $\hat{q}(t)$. At iteration $r$,
\begin{equation}
    \begin{aligned}
        \hat{q}^{r+1}(\tau) & =\hat{q}^r(\tau)\sum_{t=\max\{\tau,0\}}^{\min\{l_w-1,\tau+l_V-1\}}\frac{w(t)}{w^r(t)}V_\mathrm{PE}(t-\tau) \\
        w^r(t)              & =\sum_{\tau=\max\{t,-l_V+1\}}^{\min\{l_w-1,t-l_V+1\}} \hat{q}^r(\tau)V_\mathrm{PE}(t-\tau)
    \end{aligned}
\end{equation}
where $t\in[0,l_w-1],\tau\in[-l_V+1,l_w-1]$. $l_w$ represents the length of $\bm{w}$, and $l_V$ represents the length of $V_\mathrm{PE}$. The initial $\hat{q}^0$ could be any non-negative array whose summation is equal to that of $\bm{w}$. The two equations are two convolutions
\begin{equation}
    \begin{aligned}
        \hat{q}^{r+1}(\tau) & =\hat{q}^r(\tau)\left(\frac{\bm{w}}{\bm{w}^r}\otimes V'_\mathrm{PE}\right)(\tau+l_V-1) \\
        w^r(t)              & =(\hat{q}^r\otimes V_\mathrm{PE})(t)
    \end{aligned}
\end{equation}
where $V'_\mathrm{PE}$ is the reversed array of $V_\mathrm{PE}$.

In practice, we choose $r$ up to 2000, and the initial $\bm{z}$ is constructed from $\{\tau | \hat{q}^{2000}(\tau)>0.1\}$. If the set is empty, the corresponding waveform will be treated as zero PE, and FSMP will not analyze it.\footnote{In the numerical experiments in \zcref{sec:perf} and only when $\lambda=1$, the deconvolution method gives 4 more ``empty'' waveforms than the truth, out of 100,000 waveforms.}
The initial value of $t_0$ depends on the intensity function.
When it is unknown, the first PE time from the initial $\bm{z}$ is used.
$\hat{q}^{2000}(t)$ is used as the initial guess for the intensity function $\lambda_0\phi(t-t_0)$.

\begin{figure}
    \centering
    \includegraphics[width=\linewidth]{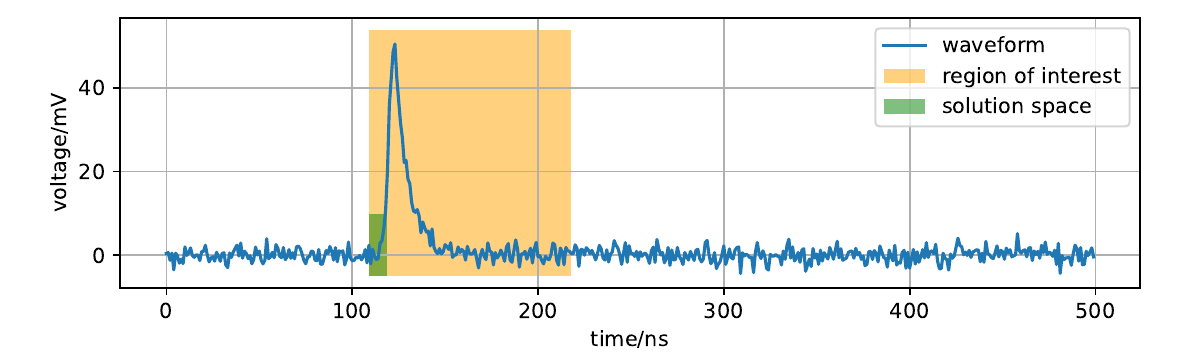}
    \caption{A simulated waveform in \zcref{sec:perf}, the region of interest and the solution space.}
    \label{fig:solution-space}
\end{figure}

\label{sec:fsmp-solution-space}
To accelerate computation, the solution space is limited by the initial PE sequence provided by deconvolution. Let the minimum and maximum PE time be $t_{\min}$ and $t_{\max}$, and the sampling rate of the waveform is \SI{1}{\giga\Hz}; the solution space $\mathcal{T}$ is $[t_{\min}-\SI{4}{\nano\second},t_{\max}+\SI{4}{\nano\second}]$. The definition range of $\bm{w}$ should be cut also to $[t_{\min}-\SI{4}{\nano\second},t_{\max}+\SI{4}{\nano\second}+l_V]$. This adjustment of \SI{4}{\nano\second} effectively expands the solution space to encompass the truth. \zcref{fig:solution-space} shows the solution space $\mathcal{T}$ and the region of interest, based on a simulated waveform in \zcref{sec:perf}.

\subsection{Sampling}
\label{sec:fsmp-mh}

Sampling of $t_0$ is done by using Metropolis-Hastings with the probability to accept the jump, or \emph{acceptance}:
\begin{equation}
    \min\left\{1, \frac{p(t'_{0,i+1}|\bm{z}_{i},\lambda_0)}{p(t_{0,i}|\bm{z}_{i},\lambda_0)}\right\}
\end{equation}
The new sample will be recorded if the jump is accepted; otherwise, the previous sample is reused. The prime in $t'_{0,i+1}$ means that the proposed value is waiting for judgment of acceptance.
$t'_{0,i+1}=t_{0,i}+\Delta t$, where $\Delta t$ follows a normal distribution $\Delta t/\si{\nano\second}\sim\mathcal{N}(0,1)$.

\label{sec:fsmp-rjmcmc}
Sampling $\bm{z}$ is done by RJMCMC.
Denote the length of $\bm{z}_i$ as $N_i$, and define the jumps: birth, death, and update in \zcref{fig:rjmcmc}. All jumps are reversible: the birth jump is the reverse of the death jump, and the update jump is the reverse of itself.

\newcommand{\action}[2]{
    \tikzmath{
        coordinate \a;
        \a = #1;
    }
    \draw[->] (\ax,\ay) -- (\ax+8cm,\ay);
    \foreach \pe/\dotted in {#2}
        {
            \draw[\dotted] (\ax+\pe,\ay) -- (\ax+\pe,\ay+1cm);
        }
    \node[] at (\ax+8.2cm,-0.5cm) {$t$};
}
\begin{figure}[!htbp]
    \centering
    \begin{subfigure}{0.32\textwidth}
        \begin{tikzpicture}[scale=0.5,line width=1pt]
            \action{(0,0)}{1cm/,3cm/,4cm/,6cm/}
            \node[] at (6cm,-0.5cm) {$t_+$};

            \draw[decorate,decoration={brace,mirror,amplitude=0.3cm}] (1cm,-0.8cm) -- node[below=0.3cm] {$\bm{z}_i$} (4cm,-0.8cm);
            \draw[decorate,decoration={brace,amplitude=0.3cm}] (1cm,1.2cm) -- node[above=0.3cm] {$\bm{z}'_{i+1}$} (6cm,1.2cm);
        \end{tikzpicture}
        \caption{}
        \label{generation}
    \end{subfigure}
    \begin{subfigure}{0.32\textwidth}
        \begin{tikzpicture}[scale=0.5,line width=1pt]
            \action{(0,0)}{1cm/,3cm/,4cm/dotted}
            \node[] at (4cm,-0.5cm) {$t_-$};

            \draw[decorate,decoration={brace,mirror,amplitude=0.3cm}] (1cm,-0.8cm) -- node[below=0.3cm] {$\bm{z}_i$} (4cm,-0.8cm);
            \draw[decorate,decoration={brace,amplitude=0.3cm}] (1cm,1.2cm) -- node[above=0.3cm] {$\bm{z}'_{i+1}$} (3cm,1.2cm);
        \end{tikzpicture}
        \caption{}
        \label{annihilation}
    \end{subfigure}
    \begin{subfigure}{0.32\textwidth}
        \begin{tikzpicture}[scale=0.5,line width=1pt]
            \action{(0,0)}{1cm/,3cm/,4cm/dotted,6cm/}
            \node[] at (4cm,-0.5cm) {$t_-$};
            \node[] at (6cm,-0.5cm) {$t_+$};

            \draw[->] plot [smooth] coordinates {(4,1.2) (5,1.4) (6,1.2)};

            \draw[decorate,decoration={brace,mirror,amplitude=0.3cm}] (1cm,-0.8cm) -- node[below=0.3cm] {$\bm{z}_i$} (4cm,-0.8cm);
            \draw[decorate,decoration={brace,amplitude=0.3cm}] (1cm,1.5cm) -- node[above=0.3cm] {$\bm{z}'_{i+1}$} (6cm,1.5cm);
        \end{tikzpicture}
        \caption{}
        \label{move}
    \end{subfigure}

    \caption{Illustrations of 3 jumps in RJMCMC.
    (\subref{generation}) Birth jump: the probability density of birth jump is $h(t_+)$, while the probability of the reverse jump is $\frac{1}{N'_{i+1}}$.
    (\subref{annihilation}) Death jump: the probability of death jump is $\frac{1}{N_i}$, while the probability density of the reverse jump is $h(t_-)$.
    (\subref{move}) Update jump: the hit time $t_-$ of one PE is updated to $t_+=t_-+\Delta t$.}
    \label{fig:rjmcmc}
\end{figure}

In the \emph{birth jump} illustrated in \zcref{generation}, a new PE $t_+$ is appended to the sequence $\bm{z}_i$. Therefore, $N'_{i+1}=N_i+1$, and $\bm{z}'_{i+1}=\bm{z}_i\cup\{t_+\}$. The proposal distribution of $t_+$ is $h(t)$. The acceptance is
\begin{equation}
    \label{eq:birth}
    \min\left\{1, \frac{p(\bm{z}'_{i+1}|\bm{w},t_{0,i+1},\lambda_0)}{p(\bm{z}_{i}|\bm{w},t_{0,i+1},\lambda_0)}\frac{\frac{1}{N'_{i+1}}}{h(t_+)}\right\}
\end{equation}

In the \emph{death jump} illustrated in \zcref{annihilation}, a PE $t_-$ is removed with equal probability from the sequence $\bm{z}_i$. Therefore, $N'_{i+1}=N_i-1$, and $\bm{z}'_{i+1}=\bm{z}_i\setminus\{t_-\}$. The acceptance is
\begin{equation}
    \label{eq:death}
    \min\left\{1, \frac{p(\bm{z}'_{i+1}|\bm{w},t_{0,i+1},\lambda_0)}{p(\bm{z}_{i}|\bm{w},t_{0,i+1},\lambda_0)}\frac{h(t_-)}{\frac{1}{N_i}}\right\}
\end{equation}

In the \emph{update jump} illustrated in \zcref{move}, a PE is moved from $t_-$ to $t_+=t_-+\Delta t$, and $\Delta t$ follows a symmetric distribution $\Delta t/\si{\nano\second}\sim\mathcal{N}(0,1)$. Therefore, $N'_{i+1}=N_i$, and $\bm{z}'_{i+1}=\bm{z}_i\setminus\{t_-\}\cup\{t_+\}$. The acceptance is
\begin{equation}
    \label{eq:update}
    \min\left\{1, \frac{p(\bm{z}'_{i+1}|\bm{w},t_{0,i+1},\lambda_0)}{p(\bm{z}_{i}|\bm{w},t_{0,i+1},\lambda_0)}\right\}
\end{equation}

The ratio terms in the acceptances are derived from \zcref{eq:pzmu,eq:pz},
\begin{equation}
    \frac{p(\bm{z}'|\bm{w},t_0,\lambda_0)}{p(\bm{z}|\bm{w},t_0,\lambda_0)}
    =\frac{p(\bm{w}|\bm{z}')p(\bm{z}'|\lambda_0,t_0)}{p(\bm{w}|\bm{z})p(\bm{z}|\lambda_0,t_0)} \\
    =\mathrm{e}^{\Delta\nu}\lambda_0^{N'-N}\frac{\phi(\bm{z}'-t_0)}{\phi(\bm{z}-t_0)}
\end{equation}
where $\Delta \nu$ is defined as
\begin{equation}
    \label{eq:delta-nu}
    \Delta\nu = \log \frac{p(\bm{w}|\bm{z}')}{p(\bm{w}|\bm{z})}
\end{equation}
\ref{sec:fsmp-fbmp} shows the calculation details, deriving from an assumption that a waveform follows a multivariate normal distribution.

In each step, at most one kind of jump is applied to a sequence. Initially, define a probability $\mathcal{J}=\frac 14$, and the probabilities for birth, death, and update jumps as $\mathcal{J},\mathcal{J},1-2\mathcal{J}$. The value of $\mathcal{J}$ is chosen from numerical experiments. We have tried values ranging from $\frac 13$ to $\frac{1}{10}$, and the chain with $\mathcal{J}=\frac 14$ converges the fastest. An exception is an empty PE sequence that could not be applied with death or update. Only the birth jump is in consideration, and the acceptance should be multiplied by the ratio of proposal $p(1\to 0)/p(0\to 1)=\mathcal{J}$. Accordingly, the acceptance of the death jump on a single PE sequence should be divided by $\mathcal{J}$.
This is to avoid proposing impossible changes and improve the acceptance of the sample chain.

In MCP-PMTs, the single PE charge follows different possible normal distributions, conditioned on the index $e$.
$\bm{z}$ should be extended as the sequence of both the time of PEs and the corresponding indices: $\bm{z}=\{(t_1,e_1), \ldots, (t_N,e_N)\}$. The MCP index of a born PE is $e_+$ and the one of a death PE is $e_-$.
An $e_+$ could generally be sampled from any proposal distribution.  We chose it to be the same as $G(e)$ for simplicity.
In the update jump, both PE time and MCP index are updated.
The discrete proposal distribution $G(e)$ of $e_+$ implies that there is a finite probability for $e_+ = e_-$ in an update.
The birth, death, and update jumps in \zcref{eq:birth,eq:death,eq:update} should be
\begin{equation}
    \label{eq:mcp}
    \begin{aligned}
         & \min\left\{1, \frac{p(\bm{z}'_{i+1}|\bm{w},t_{0,i+1},\lambda_0)}{p(\bm{z}_{i}|\bm{w},t_{0,i+1},\lambda_0)}\frac{\frac{1}{N'_{i+1}}}{h(t_+)G(e_+)}\right\} \\
         & \min\left\{1, \frac{p(\bm{z}'_{i+1}|\bm{w},t_{0,i+1},\lambda_0)}{p(\bm{z}_{i}|\bm{w},t_{0,i+1},\lambda_0)}\frac{h(t_-)G(e_-)}{\frac{1}{N_i}}\right\}      \\
         & \min\left\{1, \frac{p(\bm{z}'_{i+1}|\bm{w},t_{0,i+1},\lambda_0)}{p(\bm{z}_{i}|\bm{w},t_{0,i+1},\lambda_0)}\frac{G(e_-)}{G(e_+)}\right\}
    \end{aligned}
\end{equation}

Although the proposal distribution $h(t)$ could be any distribution covering the solution space $\mathcal{T}$, the chain will converge faster if it is proportional to the probability density $\phi(t-t_0)$. While $\phi$ is already normalized to the whole time space, it should be normalized again to $\mathcal{T}$:
\begin{equation}
    h(t)=\frac{\phi(t-t_0)}{\int_{\mathcal{T}}\phi(t-t_0)\mathrm{d}t}
\end{equation}

\subsection{Towards energy reconstruction}
\label{sec:fsmp-mu}
Visible energy $\mathcal{E}$ is the total energy of all visible photons during an event.
For the $j$th PMT, the PE count expectation $\lambda$ is proportional to the visible energy: $\lambda_j=k_j \mathcal{E}$, where $k_j$ is a function of event position and PMT quantum efficiency.
In energy reconstruction, the logarithm of the likelihood is
\begin{equation}
    \begin{aligned}
        \log L(\{\bm{w}_j\};\mathcal{E})
         & =\sum_j\log p(\bm{w}_j;\lambda_j)       \\
         & =\sum_j\log p(\bm{w}_j;k_j \mathcal{E})
    \end{aligned}
\end{equation}
The variance of MLE for energy $\hat{\mathcal{E}}$ could be calculated through the Fisher Information Matrix~(FIM)~\cite{FIM1922}:
\begin{equation}
    \begin{aligned}
        {\mathrm{Var}}^{-1}[\hat{\mathcal{E}}]
         & =\mathrm{E}\left[\sum_j \frac{\partial^2 \log p(\bm{w}_j; k_j \mathcal{E})}{\partial \mathcal{E}^2}\right] \\
         & =\sum_j \mathrm{E}\left[\frac{\partial^2 \log p(\bm{w}_j; \lambda_j)}{\partial \lambda_j^2} k_j^2\right]   \\
         & = \sum_j k_j^2 {\mathrm{Var}}^{-1}[\hat{\lambda}_j]
    \end{aligned}
\end{equation}
Where \(\mathrm{E}[\cdot]\) and \(\mathrm{Var}[\cdot]\) are the operators for expectation and variance, therefore, the improvements in $\lambda$ translate into visible energy.

The charge $Q$ of a waveform is an estimate of $\lambda$. For a dynode PMT, the charge estimator is $\hat\lambda_Q=\frac{Q}{\mu_q}$. The mean and variance are $\mathrm{E}[\hat\lambda_Q]=\lambda, \mathrm{Var}[\hat\lambda_Q]=\lambda\left(1+\frac{\sigma_q^2}{\mu_q^2}\right)$.
Despite its simplicity and popularity, this estimator is insufficient because it utilizes only the mean of the single-PE charge.

Consider the MLE of $\lambda$~\cite{Abusleme2025}:
\begin{equation}
    \begin{aligned}
        \hat\lambda_\mathrm{MLE} & =\underset{\lambda}{\arg\max}~p(Q|\lambda)               \\
                                 & =\underset{\lambda}{\arg\max}~\sum_{N}p(Q|N)p(N|\lambda)
    \end{aligned}
\end{equation}
The MLE ensures that the number of PE $N$ is an integer, and utilizes the entire charge distribution in the calculation.
$\hat\lambda_\mathrm{MLE}$ is asymptotically unbiased,\footnote{There is no reason to believe that it is unbiased in finite samples of $N$~\cite{JACQUIER2007615}, and the numerical experiments in \zcref{sec:perf} will show that it is biased.} and its variance is smaller than that of the charge estimator.
Consider the Poisson distribution $N\sim\pi(\lambda)$ and the gamma distribution $Q\sim\Gamma(Nk,\theta)$ where $\mu_q=k\theta,\sigma_q^2=k\theta^2$. The compound is a Tweedie distribution $\mathrm{Tw}(\mu,\phi,\xi)$~\cite{tweedie1984index,tweedie}:
\begin{equation}
    \lambda=\frac{1}{\phi}\frac{\mu^{2-\xi}}{2-\xi},
    k=\frac{2-\xi}{\xi-1},
    \theta=\phi(\xi-1)\mu^{\xi-1}
\end{equation}
The FIM $I$ for $\mu$, $\phi$, and $\xi$ could be calculated from the log-PDF of the Tweedie distribution.
Let $J$ be the Jacobian matrix from $\mu,\phi,\xi$ to $\lambda,k,\theta$, then the FIM for $\lambda,k,\theta$ is $I'=J^\intercal I J$~\cite{point_estimation}.
The first element of $I'$ could be calculated:
\begin{equation}
    I'_{\lambda\lambda}=\frac{k}{k+1}\frac{1}{\lambda}+J^2_{\phi\lambda}I_{\phi\phi}
\end{equation}
where $I_{\phi\phi}=-\mathrm{E}\left[\frac{\partial^2 \log f_{\mathrm{Tw}}(Q;\mu,\phi,\xi)}{\partial \phi^2}\right]>0$ because it is the MLE.
As $k,\theta$ are calibrated and fixed values in this estimator, the variance of the MLE is
\begin{equation}
    \begin{aligned}
        \mathrm{Var}[\hat\lambda_\mathrm{MLE}]
         & ={I'_{\lambda\lambda}}^{-1}                                                         \\
         & ={\left(\mathrm{Var}[\hat\lambda_Q]^{-1}+J^2_{\phi\lambda}I_{\phi\phi}\right)}^{-1} \\
         & <\mathrm{Var}[\hat\lambda_Q].
    \end{aligned}
\end{equation}
$\hat\lambda_\mathrm{MLE}$ is a more efficient estimator than $\hat\lambda_Q$, and is equivalent to $\hat{\lambda}_\mathrm{FSMP}$ without timing information. So $\hat{\lambda}_\mathrm{FSMP}$ is a better estimator than $\hat{\lambda}_Q$.


\section{Numerical experiment}
\label{sec:perf}
We conduct numerical experiments to demonstrate the energy and time resolution improvements.
The commonly used charge and first hit time methods are chosen as the baseline in order to better understand the advantages of FSMP.
We refer readers to our previous publication~\cite{xu_towards_2022} for comparison with more advanced algorithms.

The simulation chooses $\lambda$ as $0.1,5,10,\ldots,60$.
In DEAP-3600~\cite{AMAUDRUZ20191} with 255 PMTs and light yield \SI{8}{PE/\kilo\eV}, the range indicates an event with visible energy from \SI{3.2}{\keV} to \SI{1.91}{\MeV} in the detector center.
In XMASS~\cite{Abe:2008zzc} with 812 PMTs covering \SI{67}{\percent} of the surface, and \SI{4}{PE/\kilo\eV} of light yield, $\lambda=60$ is an event with visible energy \SI{18.18}{\MeV} in the detector center.
In JUNO~\cite{noauthor_juno_2022} with 17612 PMTs and light yield \SI{1350}{PE/\MeV}, it corresponds to an event with \SI{783}{\MeV} in the detector center.
Therefore, these values span experimental conditions from \si{\keV} to \si{\GeV}, and are expected to encompass energies relevant to JNE.

The \num{e8} charges for each $\lambda$ are simulated with\footnote{As JNE, our discussion context, does not use dynode PMTs, we choose a pair of typical values for Hamamatsu R12860, a kind of dynode PMT that is used in JUNO~\cite{anfimov_large_2017}.} $\mu_q=\SI{11.68}{pC},\sigma_q=\SI{3.93}{pC}$, and their MLEs are calculated.
To test the performance of FSMP, we simulate a neutrino detector with slow liquid scintillator~\cite{guo_slow_2019} that is a candidate medium of the Jinping Neutrino Experiment~\cite{xu_design_2022,luo_reconstruction_2023}.
Waveforms of \SI{500}{\nano\second} duration were simulated using \zcref{eq:waveform}, with \num{e5} samples generated for each value of $\lambda$ at a sampling rate of \SI{1}{\per\nano\second}.
For convenience, the simulated voltage values are expressed in \si{\milli\volt} as floating-point numbers, rather than integer values from an analog-to-digital converter~(ADC); this choice does not affect the validity of the results presented in this work.
The standard deviation of the white noise, $\sigma_\epsilon$, is fixed at \SI{1.55}{\milli\volt}, and the intrinsic resistance is set to \SI{50}{\ohm}.

\begin{figure}[!htbp]
    \centering
    \begin{subfigure}{0.49\linewidth}
        \includegraphics[width=\linewidth]{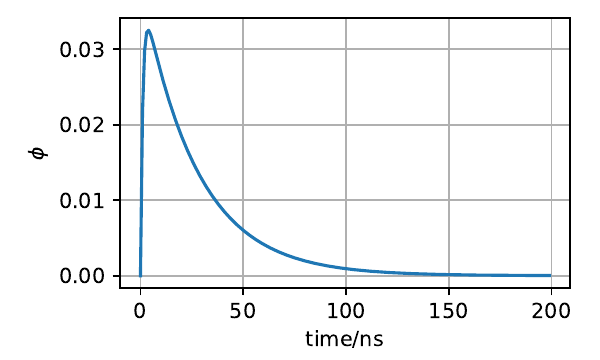}
        \caption{The probability density $\phi(t)$.}
        \label{fig:lc}
    \end{subfigure}
    \begin{subfigure}{0.49\linewidth}
        \includegraphics[width=\linewidth]{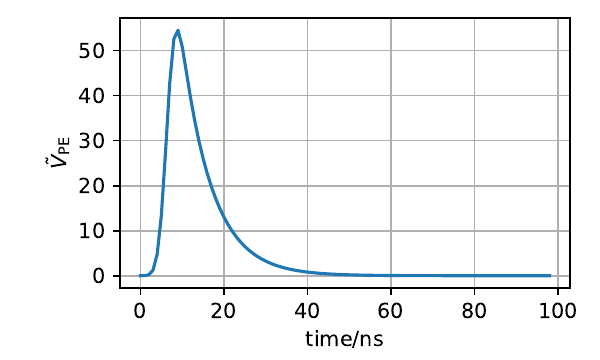}
        \caption{The SER $V_\mathrm{PE}(t)$.}
        \label{fig:ser}
    \end{subfigure}
    \caption{Figures of probability density and SER in simulation.}
\end{figure}

The probability density $\phi(t)$ in \zcref{fig:lc} is,
\begin{equation}
    \phi(t)=\frac{\tau_1+\tau_2}{\tau_2^2}\left(1-\mathrm{e}^{-\frac{t}{\tau_1}}\right)\mathrm{e}^{-\frac{t}{\tau_2}}
\end{equation}
where $\tau_1=\SI{1.16}{\nano\second}$ and $\tau_2=\SI{26.76}{\nano\second}$~\cite{guo_slow_2019} are the rise and decay time constants of the LS~\cite{LI2016303}.
There are either dynode PMTs with $\mu_q,\sigma_q$, or MCP-PMTs with parameters in \zcref{tbl:mgm,eq:mcp-multi-norm}.
The MCP charge distribution is fitted as described by Weng~et~al.~\cite{weng2024single}, and it is approximated using a Gaussian mixture model.
\zcref{fig:dynode-charge-model,fig:mcp-charge-model} plot the distribution of the charge models.
The SER in \zcref{fig:ser} is,
\begin{equation}
    V_\mathrm{PE}(t)=
    \frac{1}{2\tau}\mathrm{e}^{\frac{\sigma^2-2(t-4\sigma)\tau}{2\tau^2}}
    \left(1+\mathrm{Erf}\left(-\frac{\sigma^2-(t-4\sigma)\tau}{\sqrt{2}\sigma\tau}\right)\right)
\end{equation}
where $\sigma=\SI{1.62}{\nano\second}$, $\tau=\SI{7.2}{\nano\second}$~\cite{zhang_performance_2023}, and $\mathrm{Erf}$ is the error function.

In the simulation, we first prepare sets of waveforms with fixed PE counts \(N\) from 0 to 110. Sample \(N\) from a Poisson distribution with parameter $\lambda$ and randomly choose a waveform from the corresponding set. To sample $t_0$, a uniform distribution between ${t_0}_\mathrm{min}=\SI{100}{\nano\second}$ and ${t_0}_\mathrm{max}=\SI{200}{\nano\second}$ is chosen:
\begin{equation}
    p(t_0) = \frac{1}{{t_0}_\mathrm{max} - {t_0}_\mathrm{min}}, t_0 \in [{t_0}_\mathrm{min}, {t_0}_\mathrm{max}]
    \label{eq:t0prior}
\end{equation}

\begin{table}[!htbp]
    \centering
    \caption{Gaussian mixture parameters for MCP-PMTs.}
    \begin{tabular}{l|r|r|r}
        \hline\hline
        $e$ & $G(e)$              & $\mu_e/\si{pC}$ & $\sigma_e/\si{pC}$ \\ \hline
        0   & \SI{4.1}{\percent}  & 0.92            & 0.22               \\
        1   & \SI{51.6}{\percent} & 1.98            & 0.48               \\
        2   & \SI{18.7}{\percent} & 3.01            & 0.80               \\
        3   & \SI{19.5}{\percent} & 4.70            & 1.30               \\
        4   & \SI{6.1}{\percent}  & 6.79            & 2.04               \\
        \hline\hline
    \end{tabular}
    \label{tbl:mgm}
\end{table}

\begin{figure}[!htbp]
    \centering
    \begin{subfigure}{0.49\linewidth}
        \includegraphics[page=2,width=\linewidth]{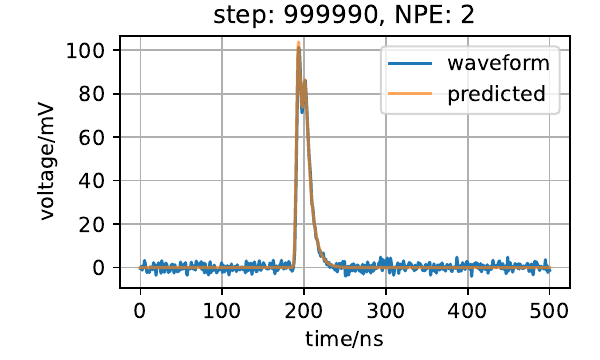}
        \caption{}
        \label{fig:fsmp-sample-a}
    \end{subfigure}
    \begin{subfigure}{0.49\linewidth}
        \includegraphics[page=2,width=\linewidth]{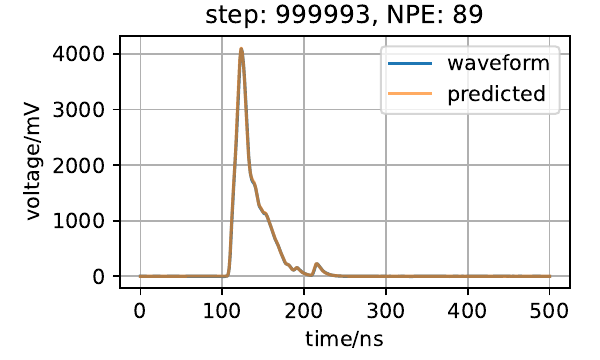}
        \caption{}
        \label{fig:fsmp-sample-b}
    \end{subfigure}
    \caption{The last sample of two Markov chains, (\subref{fig:fsmp-sample-a}) for waveform A and (\subref{fig:fsmp-sample-b}) for B. Orange lines are the predicted waveforms, very close to the original waveforms.}
    \label{fig:fsmp-sample}
\end{figure}

Two typical waveforms, one with $\lambda=1,N=2$ (waveform A) and one with $\lambda=60,N=93$ (waveform B), are drawn in \zcref{fig:fsmp-sample} to demonstrate the effectiveness of FSMP.

\subsection{Execution Speed and Precision}
\begin{figure}[!htbp]
    \begin{subfigure}{0.33\linewidth}
        \centering
        \begin{tikzpicture}[line width=1pt,minimum size=1cm,fill opacity=0.5,text opacity=1,scale=0.33]
            \foreach \x in {0,...,3}
                {
                    \tikzmath{
                        \xcenter = 6 - \x * 2;
                        \step = int(\x + 1);
                    }
                    \node[anchor=east] at (-1,\xcenter) {operation \step};
                    \foreach \ycenter in {0,2,8}
                        {
                            \draw[fill=blue] (\ycenter - 0.5,\xcenter - 0.5) rectangle (\ycenter + 0.5,\xcenter + 0.5);
                            \ifthenelse{\x > 0}{
                                \draw[->] (\ycenter,\xcenter + 1.5) -- (\ycenter,\xcenter + 0.5);
                            }{}
                        }
                    \foreach \ycenter in {2,4,8}
                        {
                            \ifthenelse{\x = 0}{
                                \draw[->] plot [smooth] coordinates {(\ycenter - 2,-0.5) (\ycenter - 1.25,-0.5) (\ycenter - 0.75,6.5) (\ycenter,6.5)};
                            }{}
                        }
                    \foreach \ycenter in {4,...,6}
                        {
                            \node[] at (\ycenter, \xcenter) {$\cdot$};
                        }
                }

            \draw[fill=blue] (0.5,-2) rectangle (1.5,-3);
            \node[anchor=west] at (1.5,-2.5) {one waveform};
        \end{tikzpicture}
        \caption{}
        \label{fig:sketch-original}
    \end{subfigure}
    \begin{subfigure}{0.33\linewidth}
        \centering
        \begin{tikzpicture}[line width=1pt,minimum size=1cm,fill opacity=0.5,text opacity=1,scale=0.33]
            \foreach \x in {0,...,3}
                {
                    \tikzmath{
                        \center = 6 - \x * 2 - 0.5;
                        \upper = \center + 0.5;
                        \downer = \center - 0.5;
                    }
                    \draw[step=1,fill=blue] (0,\upper) grid (8,\downer) rectangle (0,\upper);
                    \draw[fill=blue] (11,\upper) rectangle (12,\downer);
                    \foreach \i in {8.5,...,10.5}
                        {
                            \node[] at (\i, \center) {$\cdot$};
                        }
                    \ifthenelse{\x > 0}{
                        \draw[->] (6,\upper + 1) -- (6,\upper);
                    }{}
                }

            \draw[decorate,decoration={brace,mirror,amplitude=0.5cm}] (0.5,-1) -- (11.5,-1);

            \node[] at (6,-3) {5000 waveforms};
        \end{tikzpicture}
        \caption{}
        \label{fig:sketch-batched}
    \end{subfigure}
    \begin{subfigure}{0.33\linewidth}
        \includegraphics[width=\linewidth]{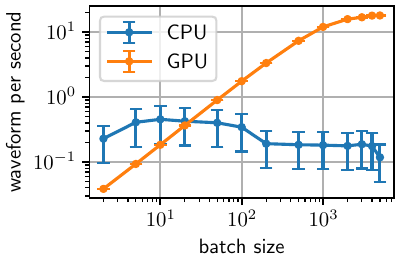}
        \caption{}
        \label{fig:prof}
    \end{subfigure}
    \caption{A comparison of (\subref{fig:sketch-original}) the original algorithm and (\subref{fig:sketch-batched}) one with a batch size of 5000. The ``operations''s are the calculations in the algorithm, and the arrows represent the execution order. The (\subref{fig:prof}) execution speed is compared on a single core of AMD EPYC\texttrademark 7742 CPU and NVIDIA\textregistered A100 GPU.}
\end{figure}

FSMP relies heavily on linear algebraic operations to compute $\Delta\nu$, which constitutes the primary performance bottleneck.
To exploit modern hardware more effectively, we reformulate the algorithm under a single-instruction multiple-data (SIMD) paradigm, enabling simultaneous analysis of multiple waveforms.
\zcref{fig:sketch-original,fig:sketch-batched} contrast the original implementation with our accelerated \emph{batched} strategy~\cite{DONGARRA2017495}, in which the number of waveforms processed concurrently is referred to as the \emph{batch size}.
In this approach, scalar, vector, and matrix quantities from different waveforms are stacked into higher-dimensional tensors with an additional \emph{batch dimension}.
The PE sequence $\bm{z} = (t_1, t_2, \ldots)$ has variable length; shorter sequences are zero-padded to form a batched matrix, and an auxiliary vector records the number of PEs $N$ for each waveform.
This scheme enables efficient FSMP implementations on both CPUs and GPUs using NumPy~\cite{harris_array_2020} and CuPy~\cite{cupy_learningsys2017}.

In \zcref{fig:prof}, the ``operation''s represent the calculations in the algorithm, \emph{e.g.}, adding two vectors. With small batch sizes, running all computation on the CPU is faster than offloading to the GPU, because data transfer between the CPU and the GPU is costly. When the batch size increases, the GPU gains performance on matrix computations up to 100 waveforms per second. The execution speed of the GPU is mostly independent of batch size.

Matrix calculation may induce float-point rounding errors.  We use \texttt{float64} on the CPU because its native instruction set is 64-bit. To better utilize the computation units~\cite{nvidia2016p100}, we choose \texttt{float32} on the GPU but with a risk of lower precision.  For comparison, every accepted step in the RJMCMC chain is recorded. We verify the GPU-driven waveform log-likelihood ratio $\Delta\nu$ in \zcref{eq:delta-nu} by the corresponding CPU program. The absolute value of the error of $\Delta\nu$ of each step is mainly within $1.0$, considered negligible.

\subsection{Convergence}
\label{sec:perf-conv}
\begin{figure}[!htbp]
    \centering
    \begin{subfigure}{0.49\linewidth}
        \includegraphics[width=\linewidth]{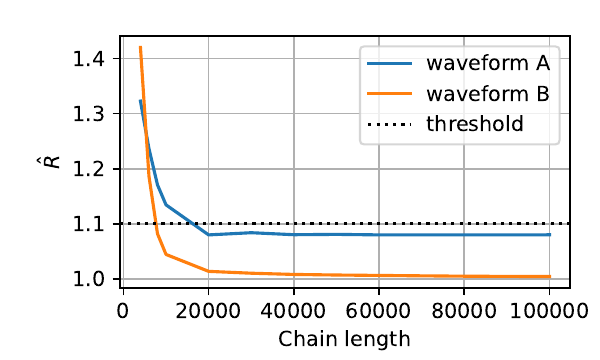}
        \caption{$t_0$ convergence of two waveforms.}
        \label{fig:t0conv}
    \end{subfigure}
    \begin{subfigure}{0.49\linewidth}
        \includegraphics[width=\linewidth]{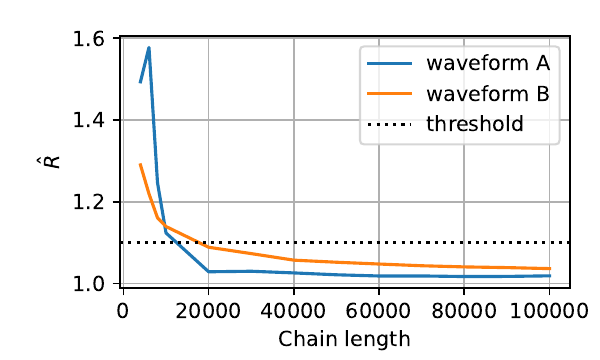}
        \caption{$N$ convergence of two waveforms.}
        \label{fig:s0conv}
    \end{subfigure}
    \begin{subfigure}{0.49\linewidth}
        \includegraphics[page=1,width=\linewidth]{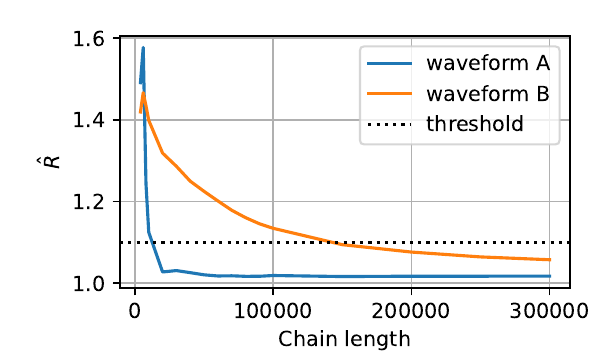}
        \caption{$\{t_i\}$ convergence of two waveforms.}
        \label{fig:sconv}
    \end{subfigure}
    \begin{subfigure}{0.49\linewidth}
        \includegraphics[page=2,width=\linewidth]{Fig7cd_gelman_w.pdf}
        \caption{$\{e_i\}$ convergence of two waveforms.}
        \label{fig:econv}
    \end{subfigure}
    \caption{The convergence of different representative scalars for waveforms in \zcref{fig:fsmp-sample}. The data point represents the \SI{90}{\percent} upper confidence limits of $\hat{R}$. All values are computed by the \texttt{gelman.diag} method in the \texttt{coda} package of the R language~\cite{coda}.}
\end{figure}

The Gelman-Rubin convergence diagnostic checks a Metropolis-Hastings Markov chain~\cite{gelman_inference_1992} by calculating an indicator $\hat{R}$ from multiple auxiliary chains with different initial conditions.  With such a combination of within-group and between-group deviations, it shows the consistency within each and among all chains.  Gelman~\cite{gelman1995bayesian} recommends regarding the chain as convergent when \(\hat{R} < 1.1\).
The initial PE sequence is randomly chosen in the solution space $\mathcal{T}$ provided in \zcref{sec:fsmp-solution-space}, ranging from 0 to 15 and 83 to 102 for waveforms A and B. For each initial $N$, there are 100 random initial $\bm{z}$.
We chose the sampled event time $t_0$ and the number of PEs $N$ as the representatives. \zcref{fig:t0conv,fig:s0conv} show the convergence of $t_0$ and $N$ of the two waveforms in \zcref{fig:fsmp-sample}.

The PE sequence \(\bm{z}\) of varied lengths, although being the most important result from FSMP, is not suitable for directly computing \(\hat{R}\).  Brooks and Gelman~\cite{brooks_general_1998} suggested several distance measures to quantify the similarity between trans-dimensional samples. We choose the Wasserstein distance~\cite{Villani2009-zw} to measure the convergence of $\bm{z}$.
As the Wasserstein distance could not handle empty sequences, a dummy PE at $t=0,e=0$ is added to all $\bm{z}$.
\zcref{fig:sconv,fig:econv} show the convergence of $\{t_i\},\{e_i\}$ of the two waveforms discussed above.
The basic trend is similar to the convergence of $t_0$ and $N$.
The slower convergence of waveform B is expected for such a large solution space that the initial conditions of the chains are diverse.
In conclusion, the chain length of \num{1.5e5} is sufficient for the chain to converge in a general, randomly initialized case.

\subsection{Bias and resolution}
\label{sec:perf-bias-res}
\begin{figure}[!htbp]
    \centering
    \begin{subfigure}{0.49\linewidth}
        \includegraphics[page=3,width=\linewidth]{Fig8_bias.pdf}
        \caption{The bias of $\hat{t}_0$ for dynode PMT.}
        \label{fig:bias_t0_dynode}
    \end{subfigure}
    \begin{subfigure}{0.49\linewidth}
        \includegraphics[page=4,width=\linewidth]{Fig8_bias.pdf}
        \caption{The resolution of $\hat{t}_0$ for dynode PMT.}
        \label{fig:res_t0_dynode}
    \end{subfigure}
    \begin{subfigure}{0.49\linewidth}
        \includegraphics[page=5,width=\linewidth]{Fig8_bias.pdf}
        \caption{The bias of $\hat{t}_0$ for MCP-PMT.}
        \label{fig:bias_t0_mcp}
    \end{subfigure}
    \begin{subfigure}{0.49\linewidth}
        \includegraphics[page=6,width=\linewidth]{Fig8_bias.pdf}
        \caption{The resolution of $\hat{t}_0$ for MCP-PMT.}
        \label{fig:res_t0_mcp}
    \end{subfigure}
    \begin{subfigure}[t]{0.49\linewidth}
        \includegraphics[page=1,width=\linewidth]{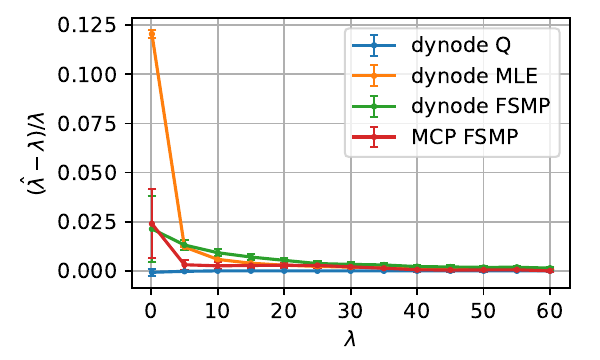}
        \caption{The relative bias of $\hat{\lambda}$.}
        \label{fig:bias_mle}
    \end{subfigure}
    \begin{subfigure}[t]{0.49\linewidth}
        \includegraphics[page=2,width=\linewidth]{Fig8_bias.pdf}
        \caption{The relative resolution of $\hat{\lambda}$. FSMP gives much better resolution. The relative resolution of the charge method is always 1.0 because it is the theoretical standard.}
        \label{fig:res_mle}
    \end{subfigure}
    \caption{The bias and resolution of $\hat{t}_0$ and $\hat{\lambda}$ from charge method $Q$ and MLE or FSMP for dynode and MCP-PMTs, with \SI{90}{\percent} confidence interval.}
\end{figure}

The estimator $\hat{t}_0$ is naturally chosen to be the average of the sampled $t_0$ chain. For comparison, the \SI{10}{\percent} rise time of the first peak~\cite{zhang_performance_2023} is chosen as a biased estimator of $t_0$.
The resolution is defined by
\begin{equation}
    \eta_t=\frac{\sqrt{\mathrm{Var}[\hat{t}_0]}}{\mathrm{E}[\hat{t}_0]}
\end{equation}
\zcref{fig:bias_t0_dynode,fig:bias_t0_mcp} show the bias of $\hat{t}_0$, while \zcref{fig:res_t0_dynode,fig:res_t0_mcp} show the resolution. FSMP is an unbiased estimator of $t_0$, and gives better resolution than the first PE time.

The energy resolution of $\hat{\lambda}$ is compared with the charge method $\hat\lambda_Q$.
The relative bias of $\hat{\lambda}$ is defined as the bias divided by the true value $(\hat{\lambda}-\lambda)/\lambda$. The resolution $\eta$~\cite{szydagis_review_2021} and relative resolution $\eta'$ of $\hat{\lambda}$ is defined as
\begin{equation}
    \eta=\frac{\sqrt{\mathrm{Var}[\hat{\lambda}]}}{\mathrm{E}[\hat{\lambda}]},\eta'=\frac{\eta}{\eta_{Q}}
\end{equation}
where $\eta_{Q}$ is the theoretical resolution of $\hat{\lambda}_Q$.

We set the intensity $\lambda_0$ to be its true values for simplicity.  In reality, $\lambda_0$ will be given by the guesses from preprocessing in \zcref{sec:fsmp-lucyddm}.  That will slightly slow down the convergence but not bias $\hat{\lambda}$.

In \zcref{fig:bias_mle,fig:res_mle}, FSMP achieves superior resolution with only a small relative bias. We anticipate that this bias will be reduced in full event reconstructions that utilize all channels simultaneously. For reference, the MLE without timing information is also shown. The improvement from MLE to dynode-FSMP and MCP-FSMP confirms that timing information also enhances energy resolution. While the advantage of FSMP is most pronounced at small $\lambda$, it still yields measurable gains at higher energies. Using $\lambda=1$ as a benchmark, FSMP outperforms the charge method by approximately \SI{5}{\percent} for dynode PMTs and \SI{10}{\percent} for MCP-PMTs in $\lambda$ estimation. In the most favorable scenario, FSMP improves the visible-energy resolution by up to \SI{10}{\percent}.


\section{Laboratory test}
\label{sec:test}
This section re-analyzes the experimental data reported by Zhang~et~al.~\cite{zhang_performance_2023} to elucidate the resolution advantages of FSMP in the presence of pile-up PEs.
In their study, they investigated the performance of a novel 8-inch MCP-PMT for JNE, conducting laboratory tests using laser light sources and CAEN V1751 1~GS/s 10-bit digitizer with effective number of bits $8.6$.
The SER is obtained from Zhang's method.
The intensity function is substituted with the deconvolution result, as explained in \zcref{sec:fsmp-lucyddm}.
$t_0$ is meaningless on this occasion, and only PE times are sampled with RJMCMC.
They could be used to measure the transition time spread~(TTS).

\begin{figure}[!htbp]
    \begin{subfigure}[t]{0.49\linewidth}
        \centering
        \includegraphics[width=\linewidth]{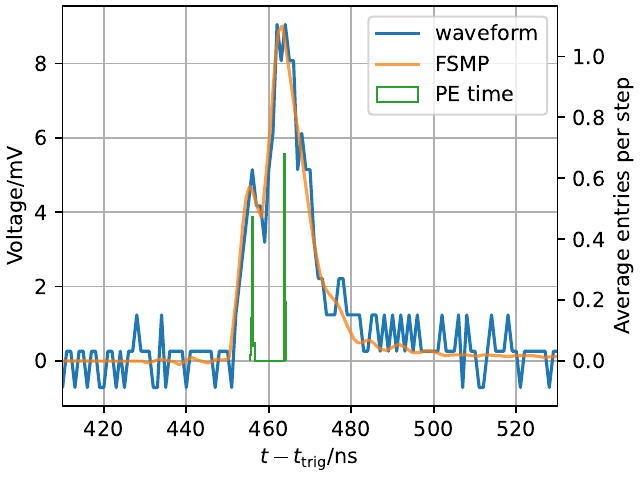}
        \caption{Average restored waveform and histogram of all sampled PE sequences of one waveform. The value of histogram represents the sampling frequency.}
        \label{fig:test-mmse}
    \end{subfigure}
    \begin{subfigure}[t]{0.49\linewidth}
        \centering
        \includegraphics[width=\linewidth,page=1]{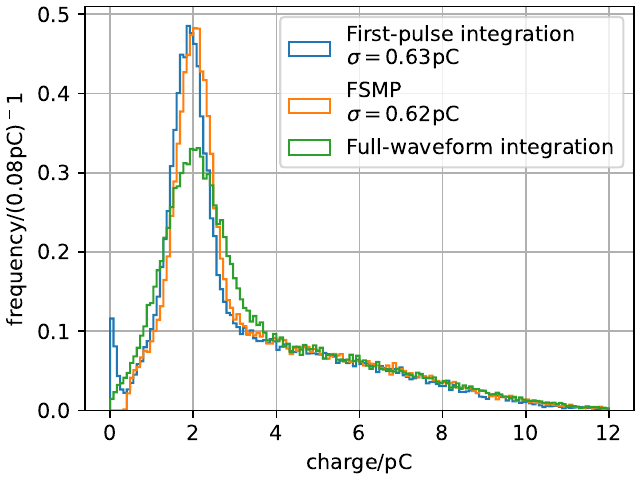}
        \caption{The charge distribution of waveforms, calculated by integration and FSMP. Only the waveform analyzed by FSMP are drawn.}
        \label{fig:test-charge}
    \end{subfigure}
    \begin{subfigure}{0.49\linewidth}
        \centering
        \includegraphics[width=\linewidth,page=1]{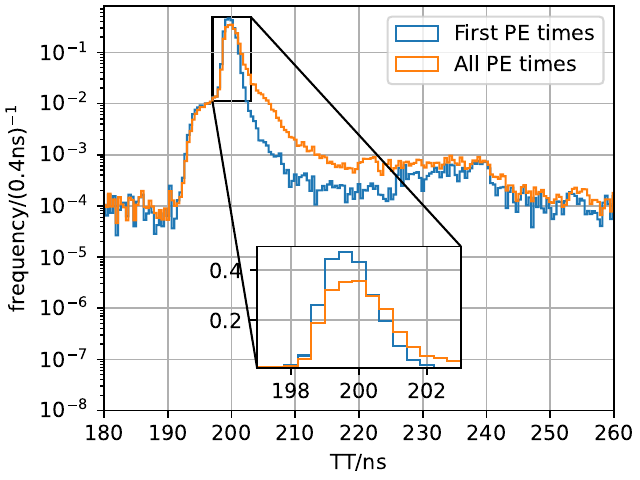}
        \caption{The TT distribution. All PE times should be used to measure the TTS.}
        \label{fig:test-tts}
    \end{subfigure}
    \begin{subfigure}{0.49\linewidth}
        \centering
        \includegraphics[width=\linewidth,page=2]{Fig9cd_tt_draw.pdf}
        \caption{The 2D distribution of charge and TT. The jumbo charge is the true secondary electrons.}
        \label{fig:test-tt}
    \end{subfigure}
    \caption{Analysis of MCP-PMT test data.}
    \label{fig:tests}
\end{figure}

We choose a sample waveform with double PEs to demonstrate the functionality of FSMP, as shown in \zcref{fig:test-mmse}. The sampled PE sequences from FSMP are convolved with the SER, restored, and averaged to the orange curve, which fits all peaks of the original waveform. The histogram of PE-time samples is drawn on the figure, averaged by sampling steps.
It indicates that the standard deviation of each PE time remains below \SI{1}{\nano\second} when the temporal separation between PEs is less than \SI{10}{\nano\second}, meeting the requirements of JNE.

\zcref{fig:test-charge} compares the charge distributions of integration and FSMP methods, in which the peak shape is fitted with a Gaussian $\mathcal{N}(Q_0,\sigma_{Q_0}^2)$~\cite{zhang_performance_2023} to calibrate the charge model and estimate the resolution. To get satisfying parameters, the integration method faces a dilemma: either to limit the integration window (the blue histogram) to reduce electronic noise but introduce bias on $Q_0$, or to integrate the total waveform (the green histogram) and sacrifice the peak resolution $\sigma_{Q_0}/Q_0$. Specifically, the length of SER $V_{\mathrm{PE}}(t)$ is \SI{152}{\nano\second}, while the integration window is \SI{85}{\nano\second}, making the charge of the integration method with a cut window \SI{4}{\percent} lower than the full charge of the waveform calculated by FSMP. The electronic noise is modeled in FSMP, giving both an unbiased charge distribution and a better peak resolution.

The transition time (TT) distribution could be calculated from the PE time samples from FSMP. \zcref{fig:test-tts} shows the distribution of TT with different definitions. The first PE times used by Zhang~et~al.~\cite{zhang_performance_2023} underestimate the TTS by ${\sim}20\%$ in terms of full width at half maxima (FWHM), with $1/20$ occupancy.  The time distribution of all PE contains a pedestal from late pulses.

\zcref{fig:test-tt} plots the distribution of charge and TT of each PE, showing structures of different physical processes.
Only the samples with $N=1$ are drawn on the figure. The main peak represents the channel-mode electrons, while the jumbo charge represents the true secondary electrons produced on the front surface. They are a little slower than the channel-mode electrons, but produce larger charges~\cite{weng2024single}. With a correlation coefficient of 0.042, the charge and TT distribution are almost independent of each other.

In summary, \zcref{fig:tests} demonstrates that FSMP gives all PE times and charges from waveforms, provides a possibility to delve into the physical process quantitatively, and provides even more evidence and knowledge about MCP-PMTs. In the future, electronics systems with higher precision might be needed to measure the time delay of true secondary electrons~\cite{Wu_2025}. We would also like to calibrate the non-linear shape adjustment of SER using linear models~\cite{Wu2024}.


\section{Conclusion}

We introduced the Fast Stochastic Matching Pursuit (FSMP), a flexible and general RJMCMC algorithm for sampling PE sequences, applicable to both dynode PMTs and MCP-PMTs. FSMP leverages both pulse-shape and amplitude information to reconstruct the complete PE sequence, thereby achieving higher precision. With GPU acceleration, FSMP attains sufficient speed to process large waveform datasets in experimental settings.

Applied to simulated waveforms, FSMP demonstrates broad applicability for enhancing energy resolution across a wide energy range, from \si{\keV} to \si{\GeV}. It also surpasses first-PE-time methods in event-time reconstruction and shows potential for distinguishing Cherenkov and scintillation photons in slow liquid scintillators. For experiments with full waveform readout, such as DEAP-3600, JNE, and JUNO, FSMP can improve visible-energy resolution by up to \SI{10}{\percent} in the most favorable scenarios.


\section{Acknowledgements}

We want to acknowledge the valuable contributions of Shengqi Chen. He provided significant help in porting the algorithm to the GPU and offered assistance with profiling. His professionalism in high-performance computing is highly appreciated. We are also grateful to Zhuojing Zhang for her inspirational guidance on RJMCMC and to Professor Zhirui Hu for the insightful discussions on the convergence of Markov chain Monte Carlo methods. We appreciate the opportunity to discuss our ideas on GPU programming provided by Tsinghua University TUNA Association.

Many thanks to Jun Weng for his patient guidance on MCP-PMT. He was one of the first FSMP users, and gave us a lot of helpful advice.
Chuang Xu and Yiqi Liu deserve our appreciation for trying FSMP with experimental data.
We are also thankful to Wentai Luo and Ye Liang for their expertise on the time properties of liquid scintillator.

This work was supported by the National Key Research and Development Program of China~(Grant no. 2023YFC3107402, 2022YFA1604704), in part by the National
Natural Science Foundation of China~(No. 12127808) and the Key
Laboratory of Particle and Radiation Imaging (Tsinghua University).
Part of the GPU computing was supported by the Center of High
Performance Computing, Tsinghua University.

\bibliographystyle{elsarticle-num-names}
\bibliography{ref}

\appendix
\section{Calculation of possibilities}

\subsection{For FSMP in dynode PMTs}
\label{sec:fsmp-fbmp}
Assume that the waveform $\bm{w}$ is a multivariate normal distribution, and the variance of electronic noise is $\sigma_\epsilon^2$. Each value of the waveform $w(t_w)$ follows a normal distribution $\mathcal{N}(\bm{U}(\mathrm{z}),\bm{\Sigma}(\mathrm{z}))$, where
\begin{equation}
    \label{eq:mgauss}
    \begin{aligned}
        U_w         & \coloneqq \sum_{k=1}^{N} \mu_q V_\mathrm{PE}(t_w-t_k)                                                              \\
        \Sigma_{wv} & \coloneqq \sum_{k=1}^{N} \sigma_q^2 V_\mathrm{PE}(t_w - t_k)V_\mathrm{PE}(t_v - t_k) +\sigma_\epsilon^2\delta_{wv} \\
                    & = \sum_{k=1}^{N} \Xi(t_w - t_k, t_v - t_k) +\sigma_\epsilon^2\delta_{wv}
    \end{aligned}
\end{equation}
Tipping~\cite{tipping_relevance_1999,tipping_sparse_2001} proves that, in this model,
\begin{equation}
    \log p(\bm{w}|\bm{z}) =
    -\frac{N_w}{2}\log (2\pi) -\frac 12\log \left|\bm{\Sigma}\right|
    - \frac{1}{2} (\bm{w} - \bm{U})^\intercal \bm{\Sigma}^{-1} (\bm{w} - \bm{U})
\end{equation}
where $N_w$ is the length of the waveform, and $\Xi$ is represented by direct product
\begin{equation}
    \Xi = \bm{a}_0 \Lambda_0 \bm{a}_0^\intercal,\ a_{0,wv}=V_\mathrm{PE}(t_w-t_v),\Lambda_{0,wv}=\sigma_q^2\delta_{wv}
\end{equation}

The update jump is a combination of the death jump at $t_-$ and the birth jump at $t_+=t_-+\Delta t$, and could be combined into one operation. For $\bm{z}'_{i+1}, t_-, t_+$ in \zcref{move}, define the waveform of PE $t_-$ as $\bm{a}_- = V_\mathrm{PE}(t_w - t_-)$. Simultaneously, define $\bm{a}_+$ as the single PE waveform of $t_+$. Combine the two waveforms into a matrix $\bm{a} = (\bm{a}_-, \bm{a}_+)$, we get
\begin{equation}
    \label{eq:11}
    \begin{aligned}
        \Delta \Sigma & = \Xi(\bm{z}') - \Xi(\bm{z}) = \bm{a} \Lambda \bm{a}^\intercal \\
        \Lambda       & \coloneqq \sigma^2_q
        \begin{bmatrix}
            -1 &   \\
               & 1 \\
        \end{bmatrix}.
    \end{aligned}
\end{equation}
For a birth jump, define $\bm{a}_-=0$; for a death jump, define $\bm{a}_+=0$. Then the 3 kinds of jumps could be unified into one formula.

RJMCMC needs $\Delta\nu$,
\begin{equation}
    \label{eq:14}
    \begin{aligned}
        \Delta\nu = \log \frac{p(\bm{w}|\bm{z}')}{p(\bm{w}|\bm{z})} \coloneqq & -\frac{1}{2}\left(\Delta T + \Delta R \right)                                                                                  \\
        \Delta T \coloneqq                                                    & \log\left(\frac{|\Sigma(\bm{z}')|}{|\Sigma(\bm{z})|}\right)                                                                    \\
        \Delta R \coloneqq                                                    & [\bm{w} - \bm{U}(\bm{z}')]^\intercal \Sigma^{-1}(\bm{z}') [\bm{w} - \bm{U}(\bm{z}')]                                           \\
                                                                              & - [\bm{w} - \bm{U}(\bm{z})]^\intercal \Sigma^{-1}(\bm{z}) [\bm{w} - \bm{U}(\bm{z})]                                            \\
        =                                                                     & (\bm{y} - \Delta \bm{U})^\intercal \Sigma^{-1}(\bm{z}') (\bm{y} - \Delta \bm{U}) - \bm{y}^\intercal \Sigma^{-1}(\bm{z}) \bm{y}
    \end{aligned}
\end{equation}
where $\bm{y} \coloneqq \bm{w} - \bm{U}(\bm{z})$. Like \zcref{eq:11},
\begin{equation}
    \label{eq:DeltaU}
    \begin{aligned}
        \Delta \bm{U} & \coloneqq \bm{U}(\bm{z}') - \bm{U}(\bm{z})
        = q(-\bm{a}_- + \bm{a}_+)
        = \bm{a}\bm{\lambda}                                       \\
        \bm{\lambda}  & \coloneqq q
        \begin{bmatrix}
            -1 \\
            1  \\
        \end{bmatrix}.
    \end{aligned}
\end{equation}
Therefore, the most important item is $\bm{\Sigma}^{-1}$. Let $\bm{c} \coloneqq \bm{\Sigma}^{-1}\bm{a}, \bm{B} \coloneqq (\Lambda^{-1} + \bm{a}^\intercal\bm{c})^{-1}$, with the Woodbury formula~\cite{woodbury_1950}
\begin{equation}
    \label{eq:12}
    \begin{aligned}
        \Sigma^{-1}(\bm{z}') & = \left(\Sigma + \bm{a} \Lambda \bm{a}^\intercal\right)^{-1}                                                         \\
                             & = \Sigma^{-1} - \Sigma^{-1}\bm{a}(\Lambda^{-1} + \bm{a}^\intercal \Sigma^{-1}\bm{a})^{-1}\bm{a}^\intercal\Sigma^{-1} \\
                             & = \Sigma^{-1} - \bm{c}\bm{B}\bm{c}^\intercal.
    \end{aligned}
\end{equation}

Calculate $\Delta R$ with \zcref{eq:14,eq:DeltaU,eq:12}:

\begin{equation}
    \label{eq:16}
    \begin{aligned}
        \Delta R & = (\bm{y} - \bm{a} \bm{\lambda})^\intercal(\Sigma^{-1} - \bm{c}\bm{B}\bm{c}^\intercal) (\bm{y} - \bm{a} \bm{\lambda}) - \bm{y}^\intercal \Sigma^{-1} \bm{y} \\
                 & = - \Upsilon^\intercal\bm{B}\Upsilon + \bm{\lambda}^\intercal\Lambda^{-1}\bm{\lambda}
    \end{aligned}
\end{equation}
where $\Upsilon \coloneqq \bm{c}^\intercal\bm{y} + \Lambda^{-1}\bm{\lambda}$.

Calculate $\Delta T$ with \zcref{eq:11,eq:12}:
\begin{equation}
    \label{eq:17}
    \begin{aligned}
        \Delta T & = \log\left(\frac{|\Sigma + \bm{a}\Lambda\bm{a}^\intercal |}{|\Sigma|}\right) \\
                 & =\log\left(|1+\bm{a}\Lambda\bm{a}^\intercal\Sigma^{-1}|\right)                \\
                 & = \log\left(|\Lambda \bm{B}^{-1}|\right)                                      \\
                 & = -\log\left(|\bm{B} \Lambda^{-1}|\right)                                     \\
    \end{aligned}
\end{equation}

With \zcref{eq:14,eq:16,eq:17}, calculate $\Delta\nu$:
\begin{equation}
    \label{eq:18}
    \Delta\nu = \frac{1}{2} (\Upsilon^\intercal\bm{B}\Upsilon - \bm{\lambda}^\intercal\Lambda^{-1}\bm{\lambda} + \log\left(| \bm{B}\Lambda^{-1}|\right))
\end{equation}

\subsection{For FSMP in MCP-PMTs}

Obviously,
\begin{equation}
    \sum_{e\in E}G(e)=1
\end{equation}
which means that $G$ is a PDF of a discrete distribution. Then the probability in \zcref{eq:pzmu} should be
\begin{equation}
    \label{eq:mcp-pzmu}
    p( \bm{z} | \lambda, t_0) \mathrm{d}\bm{z} = e^{-\lambda} \lambda^N \prod_{k=1}^N  \phi(t_k - t_0) G(e_k) \mathrm{d}t_k
\end{equation}

Considering $e_-$ and $e_+$, redefine
\begin{equation}
    \begin{aligned}
        U_w         & \coloneqq \sum_{k=1}^{N} \mu_{e_k} V_\mathrm{PE}(t_w - t_k)                                  \\
        \Sigma_{wv} & \coloneqq \sum_{k=1}^{N} \Xi_\mathrm{PE}(t_w - t_k, t_v - t_k) +\sigma_\epsilon^2\delta_{wv} \\
        \Lambda     & \coloneqq
        \begin{bmatrix}
            -\sigma^2_{e_-} &                \\
                            & \sigma^2_{e_+} \\
        \end{bmatrix}
    \end{aligned}
\end{equation}
With the same derivation in \ref{sec:fsmp-fbmp}, we can calculate $\Delta \nu$, and finally, $\frac{p(\bm{z}'|\bm{w},t_0,\lambda_0)}{p(\bm{z}|\bm{w},t_0,\lambda_0)}$.

\end{document}